\documentclass[aps,pra,floatfix,twocolumn,showpacs,superscriptaddress,reprint]{revtex4-1}
\usepackage{amsfonts}
\usepackage{mathrsfs}
\usepackage{amsmath}
\usepackage{color}
\usepackage{graphicx}
\usepackage{bm}
\usepackage{amssymb}
\usepackage{xspace}
\usepackage{epstopdf}
\usepackage{dcolumn}
\usepackage{longtable}
\usepackage{multirow}
\usepackage[colorlinks=true, letterpaper=true, pdfstartview=FitV, linkcolor=red, citecolor=blue, urlcolor=red]{hyperref}


\begin{document}
\title{
  Gap Solitons in Spin-Orbit-Coupled Bose-Einstein Condensates in Optical Lattices
       }
\author{Yongping Zhang}
\affiliation{Quantum Systems Unit, OIST Graduate University, Onna, Okinawa 904-0495, Japan}
\author{Yong Xu}
\affiliation{Department of Physics, The University of Texas at Dallas, Richardson, Texas, 75080 USA}
\author{Thomas Busch}
\affiliation{Quantum Systems Unit, OIST Graduate University, Onna, Okinawa 904-0495, Japan}

\begin{abstract}
While different ways to realize spin-orbit coupling in Bose-Einstein condensates exist, not all are currently experimentally implementable. Here we present a detailed study  of gap solitons in a Bose-Einstein condensate with experimentally realizable spin-orbit coupling and  discuss two cases relating to a spin-dependent parity symmetry.  In the parity symmetric case, two families of fundamental gap solitons in second linear energy gap are demonstrated with opposite sign of the parity, with one family having single humped densities and the other double humped ones.  In the case of broken parity symmetry, the fundamental modes manifest  spin-polarization. Both families possess an opposite sign of the spin-polarization.

\end{abstract}





\maketitle

\section{introduction}

 The interplay between spatial periodicity and non-linear dynamical evolution harbours a large number of interesting physical many body effects.  A particularly outstanding one is the existence of unique soliton solutions that reside in the linear energy gaps between the Bloch bands \cite{Lederer, Kartashov1}.  These so-called gap solitons are characterized by several remarkable features: independent of the repulsive or attractive nature of the nonlinearity, gap-solitons are always bright solitons \cite{Pelinovsky} and even though they are spatial localized, they are intrinsically related to the extended Bloch waves \cite{ Alexander, Yongping1, Bennet,Bersch}. The general properties of gap solitons have been intensively studied in many different nonlinear periodic systems, such as optical waveguide arrays and photonic lattices \cite{Lederer}, semiconductor microcavities \cite{Tanese,Santos} and also atomic Bose-Einstein condensates (BECs) in optical lattices \cite{Morsch}. The latter systems allows very good control over all available physical parameters and matter-wave gap solitons \cite{Ostrovskaya} have been observed experimentally for the first time in 2004 \cite{Eiermann}.

Another recent area of interest related to atomic BECs is the behaviour of the atoms in the presence of spin-orbit coupling  \cite{Lin,Fu, Zhang, Qu, Olson}. For neutral atoms this corresponds to a coupling between a pseudo-spin and the motional degrees of freedom and this interplay has in recent years shown to lead to intriguing new phases \cite{Wang,Wu,Ho,Sinha,Yongping0,Hu,Li} and to modified excitations and dynamics  \cite{Yongping0,Chen, Ji, Khamehchi,Danwei, Miguel}  (see \cite{Dalibard, Galitski, Goldman, Xiangfa, Zhai} for reviews). Similarly, solitonic solutions in the presence of spin-orbit coupling are predicted to show many interesting properties \cite{Merkl, Fialko, Salasnich, Xu, Achilleos1, Achilleos2,Zezyulin,Xu2,Sakaguchi}. However, as one of  the prominent features of spin-orbit-coupled system is the presence of a broken Galilean invariance \cite{Zhu}, the conventional method of finding moving solitons using stationary ones is not applicable. It is therefore a nontrivial task to find moving soliton solutions \cite{Xu, Sakaguchi}. 

The combination of spin-orbit coupling and optical lattices in BEC experiments is particularly intriguing. Spin-orbit coupling dramatically changes the Bloch spectrum and, under suitable parameters, the lowest Bloch bands can become flat  \cite{Yongping2, Scarola}.  The collective excitation possesses novel roton-like structures \cite{Jacob}.
 These new spectra are known to modify the standard Mott-superfluid transition \cite{Qian, Zhao, Zhihao} and allow for the existence of many exotic magnetic states \cite{Cai, Piraud}. Furthermore, nontrivial dynamics, such as unconventional Bloch oscillation \cite{Larson} and anomalous Anderson localization \cite{Edmonds, Zhou, Cheng}, are predicted to appear.

Recently,  Kartashov {\it et al.} studied the properties of gap solitons in a spin-orbit-coupled BEC in one and two dimensional spin-dependent optical lattices  and demonstrated  that gap solitons can be categorized according to their spin-dependent parity, time-reversal and translational symmetries \cite{Kartashov2, Kartashov3}. These studies provide a useful guideline for finding gap soliton solutions. However, the discussed systems have not yet been experimentally realized.

So far, only two experimental approaches to prepare spin-orbit-coupled BECs in lattice potentials exist. The first is a combination of Raman lasers and a radio-frequency magnetic field  \cite{Garcia}, where both the Raman beams and the radio-frequency field couple two energetic ground states of the atoms.  The momentum imparted to the atoms from the Raman beams generates artificial spin-orbit coupling, and  meanwhile dresses the radio-frequency coupling to create spin-dependent lattice potentials. For this, however, the frequency difference between the Raman beams has to be exactly the same as the frequency of the radio-frequency field, which is a significant limitation for the parameters space that can be explored. The second approach is to adiabatically load a Raman coupled BEC into a conventional optical lattice that is formed by two counter-propagating laser beams \cite{Hamner}. A frequency difference between the lattice beams creates a moving lattice and by tuning the frequency difference, the BEC can be loaded into any Bloch band in the moving frame.

Motivated by the experimental developments reported in Ref.~\cite{Hamner} and theoretical studies \cite{Kartashov2, Kartashov3}, we present in this work a systematical study of the properties of gap solitons in an  experimentally realisable spin-orbit-coupled BEC in a one dimensional optical lattice.  The properties of gap solitons existing in  the first linear gap for the repulsive interactions were first studies in Ref.~\cite{Kartashov2}
and here we mostly focus on the two families of fundamental modes in the second gap, which have a number of interesting properties. If the detuning between the Raman beams and energy states of the atoms is zero, the fundamental modes  have well-defined parity symmetry and different families of the fundamental modes take the opposite sign of the parity.  Furthermore, the density distributions of the second family has a double humped structure, which   disappears  in the presence of small Raman coupling, when the density distribution of the second family becomes similar to that of the first family. For non-zero detuning, the Hamiltonian does not have the parity symmetry and we find that the spin compositions of different fundamental mode families differ significantly. This provides a possible experimental signature to distinguish them.

This manuscript is organized as follows. In Sec.~\ref{Sec:Model} we introduce the description of our system in terms of  dimensionless Gross-Pitaevskii (GP) equations. After that, in Sec.~\ref{Sec:BlochSpectrum}, we present the linear Bloch spectrum and Bloch waves, from which the position of the linear energy gaps are identified. The parity symmetry of the Bloch waves is illustrated. In Sec.~\ref{Sec:Zero detuning} we demonstrate in detail the properties of gap solitons considering the case of  zero detuning. The case of non-zero detuning is analyzed in Sec.~\ref{Sec:Nonzero detuning} and we conclude in Sec.~\ref{Sec:conclusion}.

\section{model}
\label{Sec:Model}

Our study is fully based on the very recent experimental realization of a spin-orbit-coupled BEC by Hamner {\it et al.} \cite{Hamner}.  To create spin-orbit coupling, they select two energy levels of $^{87}$Rb atoms, which act the pseudo-spin-$1/2$, and Raman couple them with a coupling strength $\Omega$, called the Rabi frequency, and with a detuning given by $\delta$. This Raman coupling creates the spin-orbit coupling through the momentum transfer between the Raman beams and the atoms and the Hamiltonian that results from this  coupling is given by
\begin{equation}
 H_\text{soc} = \frac{p_z^2}{2m}+\gamma p_z\sigma_z +\frac{  \delta}{2}\sigma_z+\frac{ \Omega}{2}\sigma_x.
\end{equation}
Here the $\{\sigma_i\}$ are the Pauli matrices, $m$ is the atom mass,  and $p_z$ is the momentum of the atoms along the Raman beams, which counter propagate along the $z$ direction.  Spin-orbit coupling happens in the same direction and the coefficient  $\gamma=\hbar k_\text{Ram}/m$  is proportional to the wavenumber of the Raman beams $k_\text{Ram}$.

The spin-orbit-coupled BEC is then loaded into a one dimensional optical lattice by adiabatically ramping up the lattice beams, which are optically aligned with the Raman beams.
The mean-field description of the spin-orbit-coupled lattice BEC is then given by the coupled GP equations,
\begin{align}
i \hbar   \frac{\partial \Psi}{\partial t}  =  \left[   H_\text{soc}  +\text{v} \sin^2(k_\text{lat} z) \right]   \Psi
+   2\hbar \sqrt{\omega_x\omega_y}   H_\text{non} \Psi,
\end{align}
where the nonlinear terms $H_\text{non}$ originate from the atomic interactions. They are explicitly given by
\begin{equation}
H_\text{non}= \begin{pmatrix}  a_{11}   \left|  \Psi_1 \right| ^2+ a_{12}    \left|  \Psi_2 \right| ^2    & 0 \\    0&   a_{12} \left|  \Psi_1 \right| ^2+ a_{22}   \left|  \Psi_2 \right| ^2   \end{pmatrix} ,
\end{equation}
and the spinor wave-function is
$\Psi= ( \Psi_1, \Psi_2 )^T$. Here, $k_\text{lat}$ is the wavenumber of lattice beams, $\mathrm{v}$ is lattice depth which can be tuned by changing the intensity of the lattice beams, and
$\omega_x  $ and $\omega_y$ are the harmonic trap frequency in the transversal directions. We study an elongated BEC, so that the weak harmonic trap along the $z$ direction can be neglected, and we assume that the dynamics in the transverse directions is completely frozen out due to large transversal trapping frequencies.  The  $^{87}$Rb atoms are characterized by repulsive interactions and  the relevant s-wave scattering lengths are quantified by $a_{11}, a_{22}$ and $a_{12}$.  In the experiment by Hamner {\it et al.} \cite{Hamner} the difference between the s-wave scattering lengths is very small and in the following we will assume $a_{11}=a_{22}=a_{12}=a$ for simplicity.

To find gap solitons numerically, we start from the dimensionless GP equations
\begin{align}
i   \frac{\partial \Phi}{\partial t}    =  &   \left[    -\frac{1}{2}\frac{\partial^2}{\partial z^2}
-i\gamma \frac{\partial}{\partial z} \sigma_z+\frac{\delta}{2}\sigma_z+\frac{\Omega}{2}\sigma_x      \right] \Phi \notag \\
&- \frac{\mathrm{v} }{2}\cos(2z)      \Phi
+  (\left|  \Phi_1 \right| ^2   +\left|  \Phi_2 \right| ^2 )       \Phi ,
\label{TDGP}
\end{align}
where the units of energy, time and length are $2\mathrm{E_{lat}}=\hbar^2k_\text{lat}^2/m$, $\hbar/2\mathrm{E_{lat}}$ and $1/k_\text{lat}$ respectively.
The dimensionless wave-function is given by $\Phi=\Psi \sqrt{ \hbar  \sqrt{\omega_x\omega_y} a  /\mathrm{E_{lat}} }$, so that the  number of atoms $\mathrm{N}$ turns out as
\begin{equation}
\mathrm{N}=\mathrm{N}_0 \int dz  (   \left|  \Phi_1 \right| ^2   +\left|  \Phi_2 \right| ^2  ),
\end{equation}
with
\begin{equation}
\mathrm{N}_0=\mathrm{E_{lat}} / ( \hbar \sqrt{\omega_x\omega_y} k_\text{lat} a ).
\end{equation}
 For the detailed numerical calculations presented below we use the experimental parameters given in \cite{Hamner}, $k_\text{lat}=2\pi/\sqrt{2}\lambda_\text{lat}$ with $\lambda_\text{lat}=1540 nm$  and $k_\text{Ram}=2\pi/\sqrt{2} \lambda_\text{Ram}$ with $\lambda_\text{Ram}=784 nm$, which leads to a dimensionless spin-orbit coupling parameter  $\gamma=k_\text{Ram}/k_\text{lat}=1.96$. Through out this work we set $\mathrm{v}=2$ corresponding to $4\mathrm{E_{lat}}$. The detuning $\delta$ and the Rabi frequency $\Omega$ are free parameters and in the experiment they can be tuned by the changing of the frequency difference and intensity of Raman beams, respectively.

 \begin{figure*}[t]
\scalebox{1.2}{\includegraphics*{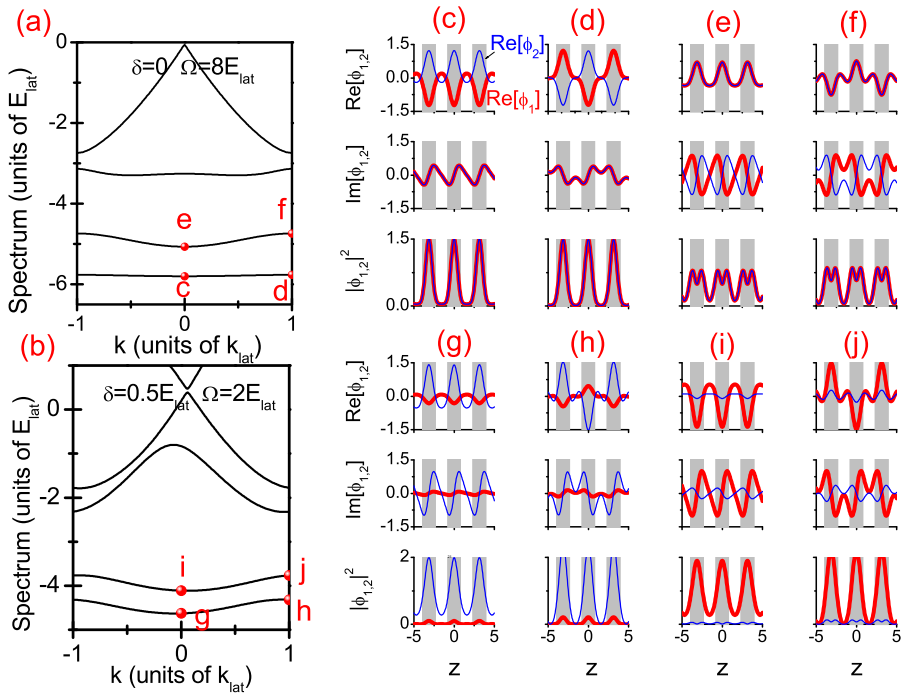}}
 \caption{(Color online) Linear Bloch spectrum and Bloch waves at the symmetry points of the spectrum. (a) The linear spectrum for zero detuning.  The Bloch waves in the lowest two bands are shown in (c) - (f), corresponding to the points indicated in (a).   (b) The linear spectrum with a finite detuning. The corresponding Bloch waves are plotted in (g) - (j).  Thick (red) lines are for $\phi_1$, while thin (blue)  lines are for $\phi_2$. The gray shadows represent $- \cos(2z) <0$. The unit of $z$ is $1/k_\text{lat}$.   Note that we only show the lowest four bands. }
 \label{fig:LinearBlochBands}
 \end{figure*}

Gap solitons are stationary solutions of Eq.~\eqref{TDGP} and have the form $\Phi(z,t)=\phi(z)\exp(-i\mu t)$, with $\mu$ being the chemical potential. The amplitude spinor $\phi(z)$ satisfies the time-independent GP equations
\begin{align}
 \mu \phi  = &  \left[ -\frac{1}{2}\frac{\partial^2}{\partial z^2}-i\gamma \frac{\partial}{\partial z} \sigma_z+\frac{\delta}{2}\sigma_z+\frac{\Omega}{2}\sigma_x  \right] \phi \notag \\
&- \frac{\mathrm{v} }{2}\cos(2z)      \phi
+  (\left|  \phi_1 \right| ^2   +\left|  \phi_2 \right| ^2 )       \phi .
\label{GP}
\end{align}
and one can see that Eq.~\eqref{GP} has a well-defined spin-dependent parity symmetry $\mathcal{P}$ if $\delta=0$
\begin{equation}
\mathcal{P}=\hat{P} \sigma_x,
\label{Parity}
\end{equation}
where $\hat{P}$ is parity operator.  This parity symmetry plays an important role in the specific form of the linear Bloch spectrum and therefore also for the properties of the gap solitons as we will show in the remainder of this paper.

Once a gap soliton solution $\phi(z)$ is found, its stability can be studied by simulating its long time behaviour using Eq.~\eqref{TDGP}. For this  we add 10\% Gaussian distributed noise to the wave function to trigger a possible instability and evolve the resulting state for 500ms. We also check the stability using the linear stability analysis by solving the standard  Bogoliubov de Gennes equations. Both, the results from nonlinear evolution and the linear stability analysis, are consistent with each other.

\section{linear Bloch spectrum and the parity symmetry of Bloch waves}
\label{Sec:BlochSpectrum}

Since gap solitons only exist in the linear energy gaps, it is important to first identify the positions of these gaps. To do this, we
use the Bloch ansatz $\phi(z)=u_{n,k}(z)\exp(ikz)$ in Eq.~\eqref{GP}, while neglecting the nonlinear terms. Here the $u_{nk}(z)$ are periodic functions, $u_{n,k}(z+\pi)=u_{n,k}(z)$, $n$ is the band index and $k$ the quasi-momentum.  The equations satisfied by the $u_{n,k}(z)$ are then
\begin{align}
\mu_{n,k} u_{n,k} =  &  -\frac{1}{2}  \left( \frac{\partial}{\partial z}+ik \right)^2 u_{n,k} - \frac{\mathrm{v} }{2}\cos(2z) u_{n,k}    \\
&  + \left[  -i\gamma  \left(  \frac{\partial}{\partial z}+ik  \right) \sigma_z \ +\frac{\delta}{2}\sigma_z+\frac{\Omega}{2}\sigma_x    \right]   u_{n,k}  \notag
\end{align}
and we calculate the linear Bloch spectrum by numerical diagonalization. 

\begin{figure}[!]
\scalebox{0.86}{\includegraphics*{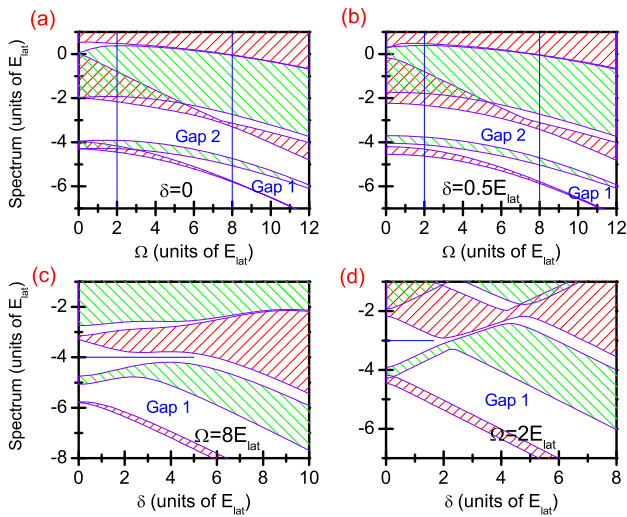}}
\caption{(Color online) Linear Bloch spectrum. Shaded regions represent the bands and white ones the gaps. (a) and (b): Spectrum as a function of the Rabi frequency $\Omega$ for $\delta=0$ and $\delta=0.5 \mathrm{E_{lat}}$, respectively.  The blue vertical lines in (a) identify the values that will be used in Figs.~\ref{fig:GapSolitonNoDetuningOmega8} and \ref{fig6}, while the ones in (b) indicate parameters that relate to results shown in Figs.~\ref{fig8} and \ref{fig9}.   (c) and (d): Spectrum as a function of the detuning $\delta$ for $\Omega=8\mathrm{E_{lat}}$ and $\Omega=2\mathrm{E_{lat}}$.  The horizontal lines in (c) and (d) mark the values relating to results shown in Figs.~\ref{fig10} and \ref{fig11}, respectively. }
\label{fig:LinearSpectrum}
\end{figure}

The application of the parity symmetry operation given in Eq.~\eqref{Parity} to  above equations results in $\mu_{n,k}=\mu_{n,-k}$, if  $\delta=0$. In fact, for these linear equations, the parity behavior is the same as the time-reversal-like symmetry $K\sigma_x$, with $K$ being the complex conjugate operator. While it might be natural to expect this symmetry in the spectrum, it is worth noting that it is broken by any non-zero detuning.  This can also be clearly seen from the two typical spectra and associated Bloch wave solutions which are shown in Fig.~\ref{fig:LinearBlochBands}: for zero detuning the spectrum is symmetric [Fig.~\ref{fig:LinearBlochBands}(a)], while for finite detuning an asymmetry is present [Fig.~\ref{fig:LinearBlochBands}(b)].

From Figs.~\ref{fig:LinearBlochBands}(c)-(f) one can see that at the points around which the spectrum is symmetric (at $k=0$ and $k=\pm 1$), the Bloch waves for $\delta=0$ possess the parity symmetry
\begin{equation}
\mathcal{P} \phi (z)=\lambda \phi (z),
\end{equation}
where the $\lambda=\pm 1$ are the eigenvalues of the parity symmetry operator $\mathcal{P}$.  This symmetry requires
\begin{equation}
\mathrm{Re}[\phi_1(z)]=\lambda \mathrm{Re}[\phi_2(-z)], \ \ \   \mathrm{Im}[\phi_1(z)]=\lambda \mathrm{Im}[\phi_2(-z)]. \notag
\end{equation}
It is very interesting to note that at these points the Bloch waves from neighboring bands have opposite parity. In Figs.~\ref{fig:LinearBlochBands}(c)-(f), the parity of the Bloch waves in the first band is $\lambda=-1$, while the parity of the Bloch waves at the second band is  $\lambda=1$. The sign of the parity of the Bloch waves depends on the sign of the Rabi frequency, $\Omega$. In the figures we show the solutions for $\Omega>0$, and for $\Omega<0$ the Bloch waves in the first band have $\lambda=1$ and the second band $\lambda=-1$. If the detuning is finite ($\delta \neq 0$), the Bloch waves become spin polarized and the polarization is determined by the sign of the detuning. Neighbouring bands have opposite spin polarization and in Fig.~\ref{fig:LinearBlochBands}(g)-(j), where we show results for $\delta>0$, the second component $\phi_2$ dominates in the first band while the first component $\phi_1$ dominates in the second band. If $\delta<0$, the results for the first and second band are inverted.

It is well-know that fundamental gap solitons  can be understood as bifurcations  from the linear Bloch waves at the symmetry points of the spectrum \cite{Pelinovsky}. Considering the intrinsic relationship between the fundamental gap solitons and the Bloch waves \cite{ Alexander, Yongping1, Bennet,Bersch}, we expect that the two features for the Bloch waves presented above are inherited by the fundamental solitons. 

 The evolution of the linear Bloch band-gap structure as a function of the different parameters is given  in Fig.~\ref{fig:LinearSpectrum}  by extracting the minima and maxima of $\mu_{n,k}$ for each band in Fig.\ref{fig:LinearBlochBands}(a)-(b).  The shaded regions represent Bloch bands while the blank areas are the linear energy gaps. From the Fig.~\ref{fig:LinearSpectrum}  the location of each gap can be easily determined.  In following sections, we will study the existence of gap solitons for the parameters indicated by the vertical and horizontal lines in Fig.~\ref{fig:LinearSpectrum}.

\section{  Gap solitons for the case of  zero detuning}
\label{Sec:Zero detuning}

\begin{figure}[tb]
\scalebox{0.85}{\includegraphics*{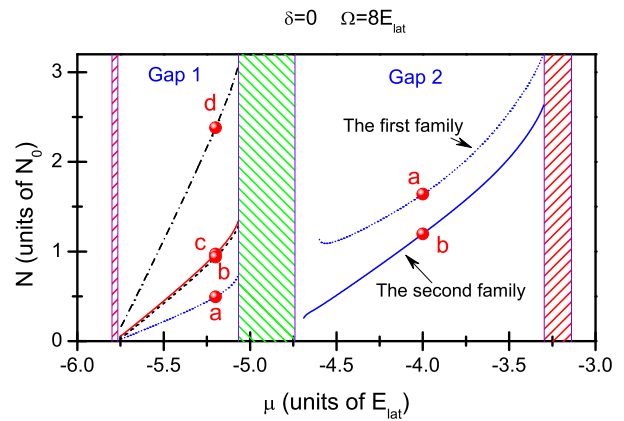}}
\caption{(Color online)  Gap solitons in the first and second gaps in the ($\mathrm{N},\mu$) plane for $\delta=0$ and $\Omega=8\mathrm{E_{lat}}$. Shaded areas correspond to the linear bands. The soliton profiles at the labeled points in the first gap are shown in Fig.~\ref{fig4} and the ones in the second gap are shown in Fig.~\ref{fig5}.  }
\label{fig:GapSolitonNoDetuningOmega8}
\end{figure}

In this section we study the gap solitons for the Rabi frequencies corresponding to the vertical lines in Fig.~\ref{fig:LinearSpectrum}(a). The gap solitons are found numerically by solving Eq.~\eqref{GP} using the Newton relaxation method.  One  general feature of gap solitons in the absence of detuning ($\delta=0$) is that they are unpolarised ($|\phi_1(z)|^2=|\phi_2(z)|^2$) and solutions for the first and second gap for $\Omega=8\mathrm{E_{lat}}$ are shown in Fig.~\ref{fig:GapSolitonNoDetuningOmega8}.

\begin{figure}[tb]
\scalebox{0.92}{\includegraphics*{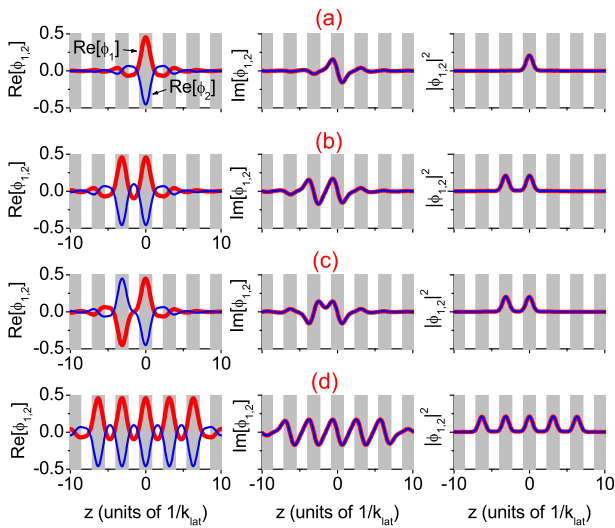}}
\caption{(Color online)  Soliton profiles at the points indicated in the first gap in Fig.~\ref{fig:GapSolitonNoDetuningOmega8}. The grayed areas correspond to $-\cos(2z) <0$. The thick (red) lines are for $\phi_1$ and  the thin (blue) lines are for $\phi_2$.  }
\label{fig4}
\end{figure}

In the first gap only one family of fundamental modes, called the first family \cite{Yongping1, Yongping3}, exists.  Their spectrum is the lowest branch (dotted line) in the first gap and they are characterized by one main density peak confined in a unit cell as shown in Fig.~\ref{fig4}(a). From Fig.~\ref{fig4}(a) it can be seen that  the first family has a parity symmetry with $\lambda=-1$. This family is expected to bifurcate from the linear Bloch wave in the first band at the Brillouin zone edge \cite{Pelinovsky} and it is very interesting to note that the parity of this family is same as the one of the corresponding linear Bloch wave (compare Fig.~\ref{fig:LinearBlochBands}(d)). The nonlinear evolution demonstrates that the whole family is dynamically stable.

\begin{figure}[tb]
\scalebox{0.25}{\includegraphics*{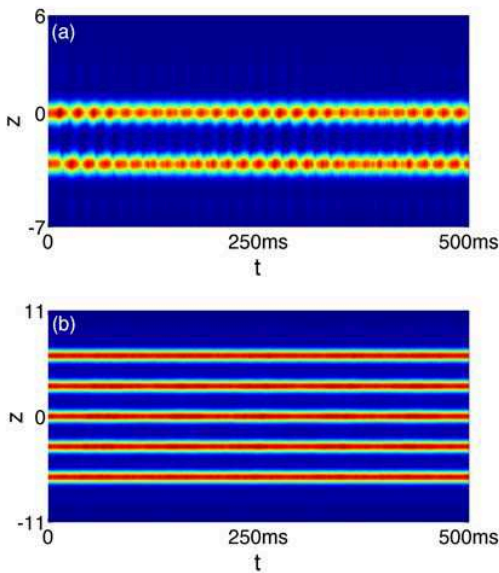}}
\caption{(Color online)   Time evolution of gap solitons in the first gap shown in Fig.~\ref{fig:GapSolitonNoDetuningOmega8}. The initial wave functions are the gap soliton solutions with 10\% Gaussian distributed noise added on. Plot (a) shows the out-phase mode of Fig.~\ref{fig4}(c) and plot (b) shows the in-phase mode shown in Fig.~\ref{fig4}(d). Only the density distribution of the first component, $|\phi_1|^2$, is shown. }
\label{fig5}
\end{figure}

\begin{figure}[tb]
\scalebox{0.92}{\includegraphics*{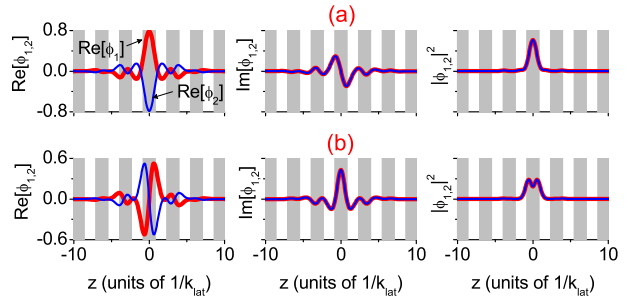}}
\caption{(Color online)  Soliton profiles at the points indicated in the second gap in Fig.~\ref{fig:GapSolitonNoDetuningOmega8}. (a) the first family of the fundamental mode and (b) the second family. }
\label{fig6}
\end{figure}

Higher-order gap solitons in the first gap can be understood as composites from the fundamental mode.  Three typical higher-order modes can be seen in Fig.~\ref{fig:GapSolitonNoDetuningOmega8} and the corresponding mode functions are shown in Fig.~\ref{fig4}(b)-(d). The two dipole modes visible in Fig.~\ref{fig:GapSolitonNoDetuningOmega8} can be seen in Fig.~\ref{fig4}(b) and (c) to correspond to a composite structure made from two fundamental modes in- and out-of-phase, respectively.  The out-of-phase dipole modes are slightly more energetic than in-phase modes in the ($\mathrm{N}, \mu$) plane. The whole branch of the out-of-phase modes are unstable, while the in-phase modes are stable except very near the linear Bloch bands. A typical unstable evolution of the out-of-phase mode triggered by the addition of an initial 10\% Gaussian distributed noise is shown in Fig.~\ref{fig5}.   Additional higher-order modes can be constructed by including more the fundamental modes.
The branch given by the dot-dash line in Fig.~\ref{fig:GapSolitonNoDetuningOmega8} in the first gap corresponds to solitons composed out of five in-phase fundamental modes and its profiles are shown in Fig.~\ref{fig4}(d). Note that the five in-phase gap solitons grow in size and are stable except very close to the linear Bloch bands. This stable evolution is shown in Fig.~\ref{fig5}.

\begin{figure}[tb]
\scalebox{0.25}{\includegraphics*{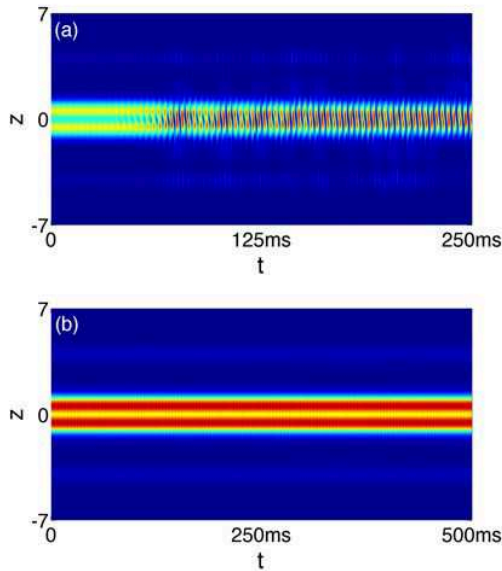}}
\caption{(Color online)   Time evolution of the second family in the second gap as shown in Fig.~\ref{fig:GapSolitonNoDetuningOmega8}. 10\% Gaussian distributed noise was added in the initial wave function to trigger potential instabilities. Plot (a) shows an unstable evolution  at $\mu=-3.75$ and plot (b) a  stable evolution  at $\mu=-3.7$. Only the density distribution of the first component, $|\phi_1|^2$, is shown. Note the different scale of $t$ in (a) and (b). }
\label{fig7}
\end{figure}

In the second gap we only consider the fundamental modes. There are two fundamental modes families, the first family (dotted line in Fig.~\ref{fig:GapSolitonNoDetuningOmega8} in the second gap) and the second family (solid line). Their profiles are shown in Fig.~\ref{fig6}.  The first family in the second gap can be considered as a continuous extension of the first family in the first gap, as they have the same parity symmetry (i.e., $\lambda=-1$) and density distribution.  The second family is particularly interesting, as they have double humped densities in the unit cell. This is very different from the case without spin-orbit coupling, where there is a node in density. Furthermore, the second family has a parity symmetry of $\lambda=1$, as can be seen from Fig.~\ref{fig6}(b). The parity of the second family is the same as the one of the linear Bloch waves in the second band in Figs.~\ref{fig:LinearBlochBands}(e)-(f). The stability analysis demonstrates that the first family is stable  except very close to the linear Bloch bands, while the second family is stable for most of the chemical potentials except a very narrow regime $-3.9<\mu <-3.74$ where it is unstable. For comparison, we show an unstable mode at $\mu=-3.75$ and a stable mode $\mu=-3.7$ in Fig.~\ref{fig7}.

\begin{figure}[t]
\scalebox{0.8}{\includegraphics*{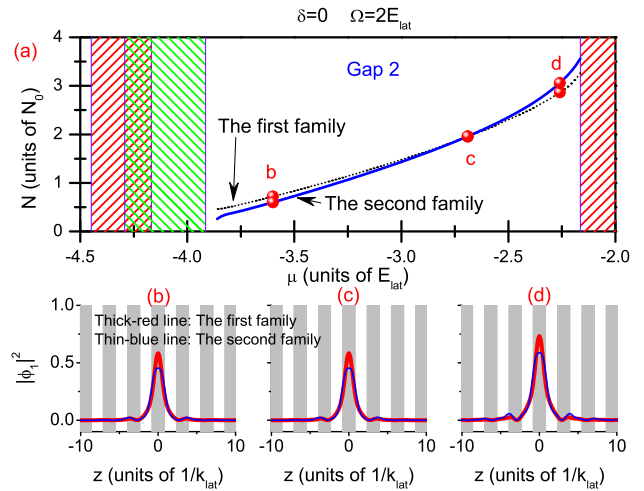}}
\caption{(Color online)  Fundamental families in the second gap for $\delta=0$ and $\Omega=2\mathrm{E_{lat}}$. Shaded areas represent the linear bands.  (a) The dotted line corresponds to the first family and the solid line to the second family.  The profiles of the labeled points are shown in (b)-(d). Since $|\phi_1(z)|^2=|\phi_2(z)|^2$, only the densities of $\phi_1$ are shown and the thick (red) lines correspond to the first family, while the thin (blue) lines correspond to the second family. }
\label{fig8}
\end{figure}

For the case of $\Omega=2\mathrm{E_{lat}}$,  the first and second linear bands overlap (see Fig.~\ref{fig:LinearSpectrum}(a)). The first gap cannot open, but the second gap is large. The spectrum of two fundamental families in this second gap is  shown in Fig.~\ref{fig8}(a) and, surprisingly,  the two families are almost degenerate in the ($\mathrm{N}, \mu$) plane and a crossover  happens around $\mu=-2.75$. The density distributions of these two families (shown in Figs.~\ref{fig8}(b)-(d)) are also very similar and only show small differences in the respective amplitudes.  Both families are characterized by one main density peak within a unit cell. However, the parity symmetry of these two families takes the opposite sign just like in the second gap of Fig.~\ref{fig:GapSolitonNoDetuningOmega8}(a), i.e., the first family has $\lambda=-1$ and the second family has $\lambda=1$. As in experiments the presence of  matter-wave gap solitons is usually probed by observing the density distribution \cite{Eiermann},  the small density difference between these two families might not be sufficient to distinguish these two families with current experimental abilities.  For chemical potentials smaller than the crossover point,  the second family is stable, while beyond the crossover point it becomes unstable. The first family is stable except close to the third linear band.

\begin{figure}[t]
\scalebox{0.78}{\includegraphics*{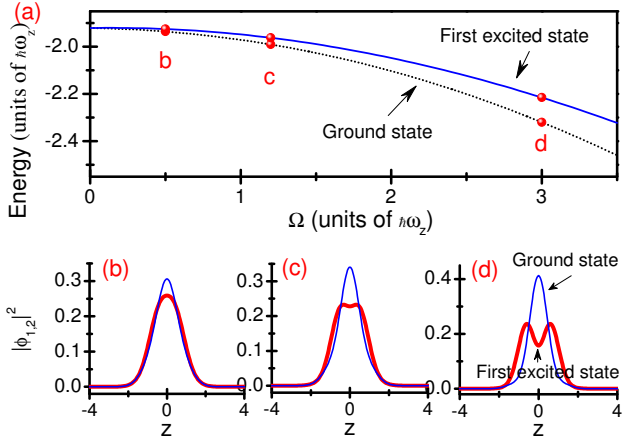}}
\caption{ (Color online)  Eigenspectrum and eigenstates of a spin-orbit-coupled single particle system in a harmonic trap with the Hamiltonian given in Eq.~\eqref{Harmonic}.  (a) Lowest two eigenenergies. The dotted line is the ground state energy while the solid line is the energy of the first excited state.  (b)-(d) Density distributions of corresponding eigenstates as indicated in (a).  Since $|\phi_1(z)|^2=|\phi_2(z)|^2$ we only show one component and the dotted lines are the densities of the ground state and solid lines are the ones of the first excited state. Here $\gamma=1.96$ and the unit of $z$ is $\sqrt{\hbar/m\omega_z}$ .  }
\label{fig9}
\end{figure}

We have seen that for a large Rabi frequency ($\Omega=8\mathrm{E_{lat}}$), the densities of the second family have a double humped structure,  while for a smaller Rabi frequency ($\Omega=2\mathrm{E_{lat}}$), the double humped structure disappears and the densities of the second family becomes single peaked. To understand this physically, we approximate the optical lattice potential of a unit cell by a harmonic oscillator potential and study in the following the energy spectrum and eigenstates of a spin-orbit-coupled single particle system in a harmonic trap. To do this, we start from corresponding Hamiltonian,
\begin{equation}
H_{\text{Har}}=\frac{p_z^2}{2m}+\gamma p_z\sigma_z +\frac{ \Omega}{2}\sigma_x+\frac{1}{2}m\omega_z^2 z^2,
\end{equation}
where $\omega_z$ is the harmonic trap frequency.  We use the harmonic oscillator basis by  replacing $p_z=i\sqrt{m\hbar \omega_z/2}(a^\dagger-a)$ and $z=\sqrt{\hbar/2m\omega_z}(a^\dagger+a)$ so that the Hamiltonian becomes,
\begin{equation}
H_{\text{Har}}= \hbar \omega_z a^\dagger a +i \gamma \sqrt{  \frac{m\hbar \omega_z}{2}  }(a^\dagger -a)\sigma_x+ \frac{\Omega }{2} \sigma_x,
\label{Harmonic}
\end{equation}
and we diagonalize it using  $\hbar \omega_z$ as the unit of energy. The ground and first excited states of this Hamiltonian are shown in Fig.~\ref{fig9}, where one can clearly see that with increasing Rabi frequency the energy splitting between these two states increases. For a small Rabi frequency,  the densities of both ground and first excited states have a single peaked structure [see Fig.~\ref{fig9}(b)], but when increasing the Rabi frequency, the density peak of the first excited state becomes flat and gradually develops a double humped profile [see Fig.~\ref{fig9}(c) and (d)]. Meanwhile the ground state keeps its single peak form.

Approximating now a periodic optical lattice potential by continuously connecting harmonic oscillator traps, the first and second families of the fundamental gap solitons in the lattice can be considered as extensions of the ground and first excited states in an individual harmonic trap respectively. Therefore some of the basic properties of the eigenstates of the harmonic trap will be passed on to the fundamental gap solitons and based on the profiles of the first excited states one can physically understand the appearance and disappearance of the double humped structure in the second family. Furthermore, we have also checked that the parity of the ground and first excited states is same as that of the first and second families respectively.   If we neglect the kinetic energy and without considering the optical lattice, our system in Eq. (\ref{GP}) becomes a nonlinear Dirac equation \cite{Cooper} and we note that the double humped density is one of the features of solitons in the nonlinear Dirac equation \cite{Cooper}.

Finally, we stress that the parity of the linear Bloch waves, the fundamental gap solitons and the eigenstates of the harmonic trap relates to the sign of the Rabi frequency. If  the sign of the Rabi frequency changes, the parity switches sign accordingly.

\begin{figure}[t]
\scalebox{0.9}{\includegraphics*{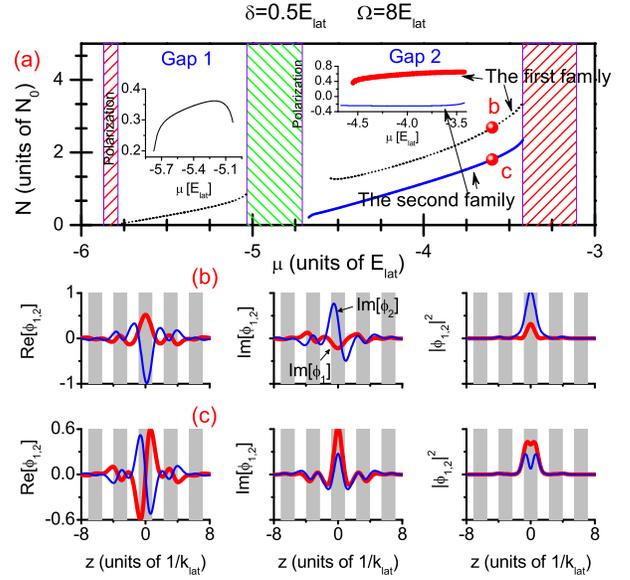}}
\caption{(Color online)  Fundamental gap solitons in the first and second gap for $\delta=0.5\mathrm{E_{lat}}$ and $\Omega=8\mathrm{E_{lat}}$.  (a) Soliton families in the ($\mathrm{N}, \mu$) plane. The dotted lines correspond to the first family and solid line to the second family. The shaded regions are the linear Bloch bands and in the embedded figures the spin polarization of corresponding families are shown. The soliton profiles at the labeled points are shown in (b) and (c), where the thick (red) lines are for $\phi_1$ and  the thin (blue) lines are for $\phi_2$. }
\label{fig10}
\end{figure}

\begin{figure}[t]
\scalebox{0.85}{\includegraphics*{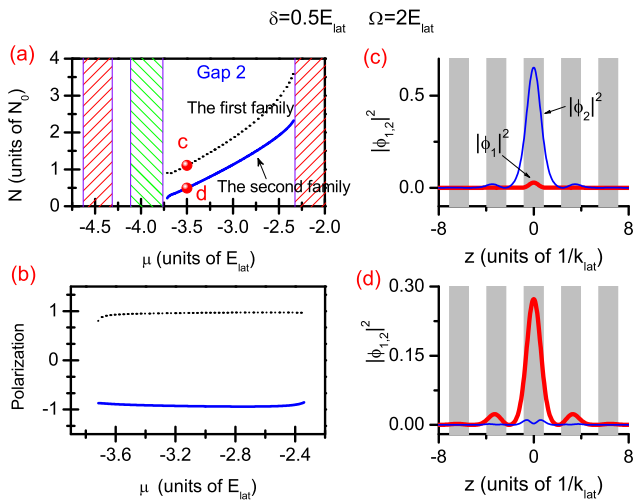}}
\caption{(Color online)  Fundamental gap solitons in the second gap for $\delta=0.5\mathrm{E_{lat}}$ and $\Omega=2\mathrm{E_{lat}}$.  (a) Soliton families and (b) spin polarization.  In both plots the dotted line corresponds to the first family and solid line to the second family.  The shaded areas represent the linear Bloch bands.  The soliton profiles of labeled points in (a) are shown in (c) and (d).  The thick (red) lines are for $\phi_1$ and  the thin (blue) lines are for $\phi_2$.  The gray shadows indicate $-\cos(2z) <0$.  }
\label{fig11}
\end{figure}

\begin{figure}[!]
\scalebox{0.9}{\includegraphics*{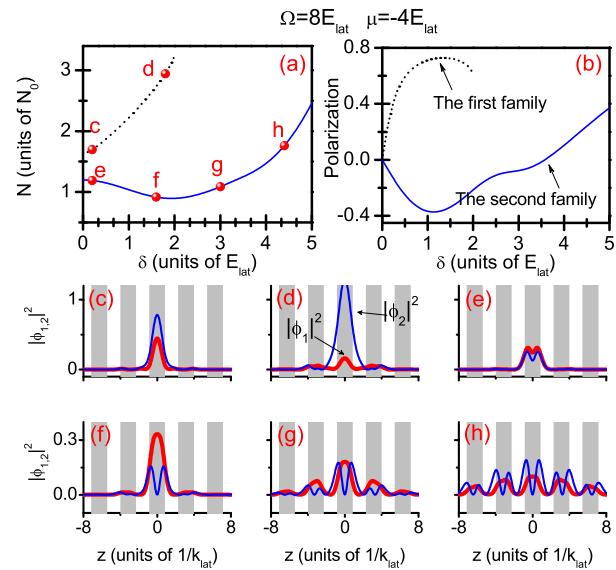}}
\caption{(Color online)  Behaviour of the first and second fundamental families in the second gap with respect to detuning for fixed chemical potential $\mu=-4\mathrm{E}_l$, i.e., along the horizontal line in Fig.~\ref{fig:LinearSpectrum}(c) with $\Omega=8\mathrm{E}_l$.   (a) Soliton families in the ($\mathrm{N}, \delta$) plane and (b) spin polarization. In (a) and (b) the dotted lines correspond to the first family and the solid lines to the second family. The density distributions of the labeled points in (a) are shown in (c)-(h). The thick (red) lines are for $\phi_1$ and  the thin (blue) lines are for $\phi_2$.  The shaded areas represent $-\cos(2z) <0$.   }
\label{fig12}
\end{figure}

\section{Gap solitons  for the case of nonzero detuning}
\label{Sec:Nonzero detuning}

The experimental preparation of a spin-orbit coupled BECs with zero detuning into the ground state is often accompanied by strong heating \cite{Lin, Hamner2}, and finite detuning is known to help to suppress this effect. In this section, we study the case of nonzero detuning in which the spin composition of the gap solitons is unbalanced and parity symmetry is absent. To quantify the spin composition we define the spin polarization as
\begin{equation}
 S= \frac{ \int dz \left [ |\phi_2(z)|^2 - |\phi_1(z)|^2 \right ] }{  \int dz \left [ |\phi_2(z)|^2 + |\phi_1(z)|^2 \right ] },
\end{equation}
which is non-zero for $\delta \neq 0$.

Gap solitons along the vertical line ($\delta=0.5\mathrm{E_{lat}}$ and $\Omega=8\mathrm{E_{lat}}$) in Fig.~\ref{fig:LinearSpectrum}(b) are shown in Fig.~\ref{fig10}.  Here we only show the fundamental modes, i.e., the first family in the first gap and the first and second families in the second gap. In the inset figures, the polarization of corresponding families is illustrated.  The polarization of the first families in the first and second gap is positive while the polarization of the second family is negative. This can be confirmed by looking at the soliton profiles in Figs.~\ref{fig10}(b) and (c).  The second component ($\phi_2$) dominates in the first family (Fig.~\ref{fig10}(b)) while the first component ($\phi_1$) is larger for the second family (Fig.~\ref{fig10}(c)).  The spin polarization of these two families coincides with that of linear Bloch waves in the first and second bands as shown in Fig.~\ref{fig:LinearBlochBands}(g)-(j).  Soliton profiles of the first family in the first gap are same as the ones in Fig.~\ref{fig10}(b). The second family still features a double humped density structure in each component, while the first family is single-peaked. The whole branch of the first family in the first gap is stable, while the first family in the second gap is stable in the regime $\mu<-4.1$, with the exception of points very close to the second linear band. Except for inside a narrow regime given by $-4.1<\mu<-3.6$, the second family is stable.

The gap solitons corresponding to the values indicated by the other vertical line ($\delta=0.5\mathrm{E_{lat}}$ and $\Omega=2\mathrm{E_{lat}}$) in Fig.~\ref{fig:LinearSpectrum}(b) are shown in Fig.~\ref{fig11}.  A narrow first gap has opened due to the small detuning, which was previously closed in the linear spectrum for zero detuning.  Since the first gap is narrow, it is very difficult to find gap solitons there and we concentrate on studying the fundamental modes in the second gap in which two families can be found. These have the same properties as the ones in the second gap of Fig.~\ref{fig10}(a), however, compared to the case of zero detuning shown in Fig.~\ref{fig8}, these two families do not overlap in the ($\mathrm{N},\mu$) plane (see Fig~.\ref{fig11}(a)). Furthermore, the density distributions of the second family are different  compared to the case of zero detuning: the component $\phi_1$ has a single main peak while the component $\phi_2$ becomes double peaked.  The stability calculations show that the whole second family is stable, while the first family is unstable for chemical potentials $\mu >-2.75$.

\begin{figure}[!]
\scalebox{0.78}{\includegraphics*{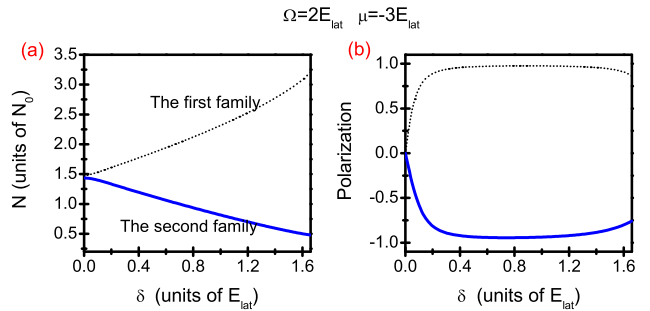}}
\caption{(Color online)  Behaviour of the first and second fundamental families in the second gap with respect to the detuning for fixed chemical potential $\mu=-3\mathrm{E}_l$ along the horizontal line in Fig.~\ref{fig:LinearSpectrum}(d).  Here $\Omega=2\mathrm{E}_l$.  }
\label{fig13}
\end{figure}

To further characterize the role of the detuning on the fundamental modes, we fix the chemical potential and study the existence of fundamental modes in the second gap with respect to a change in detuning for values indicated by the horizontal lines in Figs.~\ref{fig:LinearSpectrum}(c) and (d).  The results are shown in Figs.~\ref{fig12} and \ref{fig13}.  For $\Omega=8\mathrm{E_{lat}}$ [Fig.~\ref{fig12}], the first family (dotted line in Fig.~\ref{fig12}(a)) only exist for small detuning.  With increasing values, the number of atoms in the first family increases, while it initially decreases and only later increases for the second family.  The polarization of the second family also shows anomalous behavior for the large detuning regime. For large detuning the size of the second gap becomes very small [see Fig.~\ref{fig:LinearSpectrum}],  and the soliton profiles of the second family become less localized as shown in Figs.~\ref{fig12}(g) and (h).  For small detuning both families are stable, while they become unstable beyond the first extremum (around $\delta=1\mathrm{E_{lat}}$) in their polarization curve.

For $\Omega=2\mathrm{E_{lat}}$, as shown in Fig.~\ref{fig13}, small values of the detuning can be seen to increase the spin polarization for the first and second families, while for large detuning both values saturate. Both family are stable except for detuning  $\delta >1.1\mathrm{E_{lat}}$ where the curvature of the polarization graph is changed.

\section{conclusion}
\label{Sec:conclusion}

We have studied the  properties of gap solitons in an experimentally available spin-orbit-coupled setup. When the detuning between the Raman beams and ground energy levels of the atoms is zero, parity symmetry plays important role.  We have revealed that the two families of fundamental gap solitons in the second linear energy gap have opposite signs of parity, and the sign relates to the sign of Rabi frequency.  The first family is characterized by a single peaked density while the second family has a doubly humped density structure. However, the double humped structure disappears for small values of the Rabi frequency.  An intuitive picture, that approximates the lattice potential over a unit cell by a harmonic trap, was used to physically understand the evolution of the double humps.  When the detuning is non-zero, fundamental gap solitons become spin-polarized. The two families then have opposite polarization and the sign of polarization depends on the sign of the detuning.  The parity symmetry and spin-polarization of these two families of fundamental gap solitons correlate with that of linear Bloch waves at the symmetric points of the spectrum.

Finally, we would like to mention that beside the parity symmetry Eq.~(\ref{GP}) also has a time-reversal-like symmetry $K\sigma_x$ if $\delta=0$. 
If $\phi_0$ is a solution of this equation, then the wave-functions $\exp(i\psi)\phi_0$ with arbitrary global phase $\psi$ are also solutions and the uncertainty of the global phase makes a classification of solitons according to the time-reversal symmetry not straightforward \cite{Kartashov2}. We have therefore not discussed this case here. In addition, Ref.\cite{Kartashov2} recently explored a further type of spin-dependent translational symmetry, which originates from spin-dependent optical lattices and which is also absent in our case.

We acknowledge useful conversation with  Avadh Saxena. This work was supported by Okinawa Institute of Science and Technology Graduate University.

\bibliographystyle{apsrev4-1}

\end{document}